\documentclass[aps,prl,superscriptaddress]{revtex4}
\usepackage{mathrsfs}
\usepackage{graphicx}
\usepackage{amsfonts}
\usepackage{amsmath}
\usepackage{amssymb}
\usepackage[hypertex]{hyperref}

\begin{document}

\title{Test of Einstein-Podolsky-Rosen Steering Based on the All-Versus-Nothing Proof}

\author{Chunfeng~Wu\footnote{Correspondence to:
chunfeng\_wu@sutd.edu.sg}} \affiliation{Centre for Quantum
Technologies, National University of Singapore, 3 Science Drive 2,
Singapore 117543} \affiliation{Pillar of Engineering Product
Development, Singapore University of Technology and Design, 20 Dover
Drive, Singapore 138682}

\author{Jing-Ling~Chen\footnote{Correspondence to:
cqtchenj@nus.edu.sg}} \affiliation{Centre for Quantum Technologies,
National University of Singapore, 3 Science Drive 2, Singapore
117543} \affiliation{Theoretical Physics Division, Chern Institute
of Mathematics, Nankai University, Tianjin 300071, People's Republic
of China}

\author{Xiang-Jun~Ye}
\affiliation{Centre for Quantum Technologies, National
University of Singapore, 3 Science Drive 2, Singapore 117543}
\affiliation{Theoretical Physics Division, Chern Institute of
Mathematics, Nankai University, Tianjin 300071, People's Republic of
China}

\author{Hong-Yi~Su}
\affiliation{Centre for Quantum Technologies, National
University of Singapore, 3 Science Drive 2, Singapore 117543}
\affiliation{Theoretical Physics Division, Chern Institute of
Mathematics, Nankai University, Tianjin 300071, People's Republic of
China}

\author{Dong-Ling~Deng}
\affiliation{Department of Physics and Michigan Center for
Theoretical Physics, University of Michigan, Ann Arbor, Michigan
48109, USA}

\author{Zhenghan~Wang}
\affiliation{Microsoft Research,
Station Q, University of California, Santa Barbara, CA 93106, USA}

\author{C. H. Oh\footnote{Correspondence to:
 phyohch@nus.edu.sg}}
\affiliation{Centre for Quantum Technologies, National
University of Singapore, 3 Science Drive 2, Singapore 117543}
\affiliation{Department of Physics, National University of
Singapore, 2 Science Drive 3, Singapore 117551}

\date{\today}

\maketitle

\textbf{In comparison with entanglement and Bell nonlocality,
Einstein-Podolsky-Rosen steering is a newly emerged research topic
and in its incipient stage. Although Einstein-Podolsky-Rosen
steering has been explored via violations of steering inequalities
both theoretically and experimentally, the known inequalities in the
literatures are far from well-developed. As a result, it is not yet
possible to observe Einstein-Podolsky-Rosen steering for some steerable mixed
states. Recently, a simple approach was presented to identify
Einstein-Podolsky-Rosen steering based on all-versus-nothing
argument, offering a strong condition to witness the steerability of
a family of two-qubit (pure or mixed) entangled states. In this
work, we show that the all-versus-nothing proof of
Einstein-Podolsky-Rosen steering can be tested by measuring the
projective probabilities. Through the bound of probabilities imposed
by local-hidden-state model, the proposed test shows that steering
can be detected by the all-versus-nothing argument experimentally
even in the presence of imprecision and errors. Our test can be
implemented in many physical systems and we discuss the possible
realizations of our scheme with non-Abelian anyons and trapped
ions.}

\vspace{8pt}
\noindent In 1935, Einstein, Podolsky, and Rosen (EPR)
questioned the completeness of quantum mechanics (QM) based on local
realism~\cite{1935Einstein}. Many efforts have been devoted to a
deeper understanding of QM in the form of three types of quantum
nonlocalities: quantum entanglement, EPR steering, and Bell
nonlocality~\cite{WJD07}. Within the hierarchy of nonlocalities, the
set of EPR steerable states is a subset of entangled states and a
superset of Bell nonlocal states. Quantum entanglement and Bell
nonlocality have attained flourishing developments since 1964.
However, EPR steering is a newly emerged research topic and, to
date, is far from being completely understood. Steering inequalities
for EPR steering are the analog of Bell inequalities for Bell
nonlocality. Their violations, predicted by quantum mechanics,
reveal EPR steering. Such a violation rules out the existence of a
local-hidden-state (LHS) model, the same way the violation of a Bell
inequality rules out the existence of a local-hidden-variable (LHV)
model. In comparison to the development of Bell nonlocality, the
research on EPR steering is in its developing stages, even though
Schr{\" o}dinger discussed the concept in 1935~\cite{Schrsteering}.
A reason for this is the absence of a rigorous formulation of the
concept of EPR steering, which did not appear until the work of
Wiseman, Jones, and Doherty~\cite{WJD07} in 2007. Indeed, EPR
steering answers a question of fundamental quantum physics as well
as opens new possibilities for quantum communication, thus it has
inspired some recent research in quantum information
theory~\cite{SQKD}.

For a pure entangled state shared by two separated observers Alice
and Bob, Bob's qubit can be ``steered" into different states
although Alice has no access to the qubit. Schr\"{o}dinger adopted
the word \emph{steering}" to describe this type of nonlocality. This
means that Alice has the ability to remotely prepare Bob's particle
in different states by measuring her particle using different settings,
and here we use $\tilde{\rho}^{A}_a$ to denote the conditional state Bob gets if Alice measures her particle with measurement $\hat{A}$ and obtains result $a$.
While Bob suspects that Alice may send
him some non-entangled particles and fabricate the results based her
knowledge of LHS. If Bob's system admits a LHS model $\{ \wp_{\xi} \rho_{\xi} \}$, where $\rho_{\xi}$'s are states that Bob does not know (but Alice knows),
and $\wp_{\xi}>0$ is the probability of $\rho_{\xi}$, then Alice could attempt to fabricate the results
using her knowledge of $\xi$, in other words, $\tilde{\rho}^{A}_a=\sum_{\xi} \wp(a|\hat{A},\xi) \wp_{\xi} \rho_{\xi}$, with $\sum_{a} \wp(a|\hat{A},\xi)=1$.
If Bob finds there is a LHS model which can
describe his conditional states after he asks Alice to perform the
measurement on her particle, then he is not convinced the existence
of EPR steering.

Very recently many results have been achieved to show violations of
steering inequalities both theoretically and experimentally, thus
rendering LHS model untenable
\cite{Reid1,Reid2,stin,NP2010,NC2012,He11,Loopholefree}. However,
the existed steering inequalities in the literatures are far from
well-developed, and therefore it is not yet possible to observe EPR
steering for some steerable mixed states~\cite{EPR-us}. Another elegant
approach to explore the contradiction between QM and LHS model is
the \emph{all-versus-nothing} (AVN) proof of the existence of EPR
steering. This can be considered as the steering analog of
Greenberger-Horne-Zeilinger (GHZ) argument without inequalities for
Bell nonlocality~\cite{GHZ89}. Currently such an AVN proof for EPR
steering has been shown to be a strong condition to witness the
steerability of a family of two-qubit (pure or mixed) entangled
states and have the ability of detecting asymmetric
steering~\cite{EPR-us}. This also offers an effective way to detect
EPR steering for two qubits experimentally.

In this work, we investigate the test of EPR steering according to
its AVN argument and demonstrate directly the contradiction between
LHS model and QM. We show that by observing projective
probabilities, the existence of steering can be verified by defining
a probability bound imposed by LHS model. Our test is the first one
proposed to detect EPR steering based on the AVN proof and it is
suitable for all the two-qubit entangled states specified in
Ref.~\cite{EPR-us}, both pure and mixed. The possible implementation
of our test is discussed by using non-Abelian Fibonacci anyons and
trapped ions, but it is not limited to these systems. Our test is
also applicable to many other physical systems, such as photons,
atoms as well as superconductors, etc. In a system of non-Abelian Fibonacci
anyons, each logical qubit is encoded into triplet of Fibonacci
anyons and the corresponding operations are carried out by braiding
the anyons. As braids are performed by taking an anyon either around
another or not, which will not cause small errors from slight
imprecisions in the way that anyons are moved. Therefore, the test
is fault-tolerant to errors and offers high experimental precision.
In an ion-trap experiment, present experimental achievements on
high-fidelity state initialization, quantum gates and state readout
make our scheme of detecting steering possibly testable.


\vspace{8pt}

\noindent{\bf Results}

\noindent First let Alice and Bob share
a pure entangled state
$|\Psi\rangle_{AB}=\cos\theta|00\rangle_{AB}+\sin\theta|11\rangle_{AB}$.
In the steering scenario, Alice adopts the following settings:
$\hat{A} \in \{ \mathcal {P}^{\hat{z}}_a, \mathcal {P}^{\hat{x}}_a
\}$, where $\mathcal {P}^{\hat{z}}_a$ and $\mathcal {P}^{\hat{x}}_a$
denote Alice's projective measurements in $\hat{z}$- and
$\hat{x}$-directions, and $a$ (with $a=0, 1$) is measurement
result. After Alice's measurements, Bob's conditional states become
\begin{eqnarray}\label{crhoz}
\tilde{\rho}^{\hat{z},0}_B&=&\cos^{2}\theta \; |0\rangle_B \langle 0 |,\nonumber\\
\tilde{\rho}^{\hat{z},1}_B&=&\sin^{2}\theta  \;|1\rangle_B \langle 1 |,\nonumber\\
\tilde{\rho}^{\hat{x},0}_B&=&\frac{1}{2} |\psi\rangle_B \langle \psi|,\nonumber\\
\tilde{\rho}^{\hat{x},1}_B&=&\frac{1}{2} | \varphi \rangle_B \langle
\varphi|,
\end{eqnarray}
where $\tilde{\rho}^{\hat{A},a}_B$ describes Bob's state after Alice
performs measurement $\hat{A}$ and obtains result $a$, and $|\psi
\rangle_B= \cos \theta |0\rangle_B + \sin \theta |1\rangle_B$,
$|\varphi \rangle_B= \cos \theta |0\rangle_B - \sin \theta
|1\rangle_B$. If there exists a LHS model can fake the results
(\ref{crhoz}), i.e., there exists a suitable ensemble $\{ \wp_{\xi}
\rho_{\xi} \}$ and a stochastic map $\wp(a|\hat{A},\xi)$ satisfying
$\tilde{\rho}^{\hat{A},a}_B=\sum_\xi \wp(a|\hat{A},\xi) \wp_{\xi}
\rho_{\xi}$, then Bob is not convinced that Alice can \emph{steer}
his conditional states. Otherwise the LHS model contradicts with QM.

According to the AVN proof \cite{EPR-us}, the entangled state
$|\Psi\rangle_{AB}$ cannot be described by any LHS model except $\theta=0$ or $\pi/2$. The incisive contradiction between QM and LHS
model is due to different predicted projective probabilities as stated in the following. For QM, Bob obtains zero probabilities
after he performs some appropriate projective measurements on his
qubit
\begin{eqnarray}
P^{\rm QM}_{1}&=& {\rm Tr}[ |1\rangle_B \langle 1| \tilde{\rho}^{\hat{z},0}_B] = 0  ,\nonumber \\
P^{\rm QM}_{2}&=& {\rm Tr}[ |0\rangle_B \langle 0| \tilde{\rho}^{\hat{z},1}_B] = 0  ,\nonumber \\
P^{\rm QM}_{3}&=& {\rm Tr}[ |\psi^\perp \rangle_B \langle \psi^\perp | \tilde{\rho}^{\hat{x},0}_B] = 0  , \nonumber\\
P^{\rm QM}_{4}&=& {\rm Tr}[ |\varphi^\perp \rangle_B \langle
\varphi^\perp | \tilde{\rho}^{\hat{x},1}_B] = 0,\label{cF}
\end{eqnarray}
where
$|\psi^\perp\rangle_B=\sin\theta|0\rangle_B-\cos\theta|1\rangle_B$
and
$|\varphi^\perp\rangle_B=\sin\theta|0\rangle_B+\cos\theta|1\rangle_B$
are orthogonal to $|\psi\rangle_B$ and $|\varphi\rangle_B$,
respectively. However, for a LHS model, it predicts the corresponding probabilities as follows,
\begin{eqnarray}
&&P^{\rm LHS}_{1}={\rm Tr}[|1\rangle_B \langle 1 | \sum_\xi \wp(a=0|\hat{z},\xi) \wp_{\xi} \rho_{\xi} ], \nonumber\\
&&P^{\rm LHS}_{2}={\rm Tr}[|0\rangle_B \langle 0 | \sum_\xi \wp(a=1|\hat{z},\xi) \wp_{\xi} \rho_{\xi} ],\nonumber\\
&&P^{\rm LHS}_{3}={\rm Tr}[|\psi^\perp\rangle_B \langle \psi^\perp | \sum_\xi \wp(a=0|\hat{x},\xi) \wp_{\xi} \rho_{\xi} ], \nonumber\\
&&P^{\rm LHS}_{4}={\rm Tr}[|\varphi^\perp\rangle_B \langle
\varphi^\perp | \sum_\xi \wp(a=1|\hat{x},\xi) \wp_{\xi} \rho_{\xi}
].\label{qqcF}
\end{eqnarray}
From the AVN proof \cite{EPR-us}, we know that the state $|\Psi\rangle_{AB}$ possesses EPR steering if $\theta\neq 0$ or $\pi/2$, and this tells us there exists no LHS model of the state such that $\tilde{\rho}^{A}_a=\sum_{\xi} \wp(a|\hat{A},\xi) \wp_{\xi} \rho_{\xi}$. When $\theta=0$ or $\pi/2$, the state is separable, and hence it is possible to find a LHS model to describe it. Therefore, we know that the probabilities (\ref{qqcF}) cannot be zero simultaneously except $\theta=0$ or $\pi/2$.

%

\begin{figure}
\includegraphics[width=100mm]{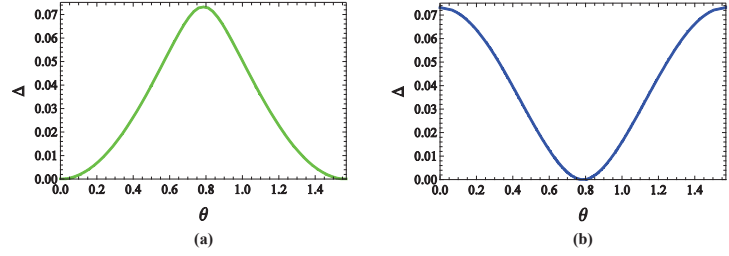}\\
\caption{ \textbf{Numerical results of the bound $\Delta$ imposed by
the optimal LHS model versus $\theta$}. For (a) $|\Psi\rangle_{AB}$
and (b) $\rho_{AB}$.}\label{LHS1}
\end{figure}

In an ideal test for EPR steering, after Alice performs projective measurement on her qubit of the state
$|\Psi\rangle_{AB}$, Bob then measures the probabilities by
projecting the states $|0\rangle_B$, $|1\rangle_B$,
$|\psi^\perp\rangle_B$ and $|\varphi^\perp\rangle_B$ on his qubits.
If he finds the four probabilities $P^B_{i}$ ($i=1, 2, 3, 4$)
are all zero, then EPR steering is demonstrated. Nevertheless, in
real experiments (Exp), measurement results are inevitably affected
by experimental precision and errors. It is possible that the
probabilities obtained experimentally may deviate from the
theoretical values slightly, i.e., $P^{\rm Exp}_{i}=P^{\rm
QM}_i+\varepsilon_i$ (here $\varepsilon_i$ are small numbers caused
by errors). We then investigate how close a LHS model could be to
simulate Eq.~(\ref{cF}). We have shown that for the state
$|\Psi\rangle_{AB}$ the probabilities $P^{\rm LHS}_{i}=0 \;(i
=1,2,3,4)$ only if the state shared by Alice and Bob is not
steerable, otherwise, some of $P^{\rm LHS}_{i}$ cannot be zero. For
the sake of simplicity, consider the inevitable errors
$\varepsilon_i=\varepsilon$ for all $i$, we can detect steerability
if $P^{\rm LHS}_{i} > \varepsilon$ for some $i $. Therefore, the
optimal LHS model for this experiment is the one making $P^{\rm
LHS}_{i}$ approach to $P^{\rm QM}_{i}$ as closely as possible for
all $i $. We define
\begin{eqnarray}
\Delta= \min_{\rm LHS} \{ \max_{i\in\{1,2,3,4\}} \{ |P^{\rm
LHS}_{i}-P^{\rm QM}_{i}|
 \}  \}, \label{qqDT}
\end{eqnarray}
where $\Delta$ describes the bound of probabilities imposed by the
optimal LHS model. In our test, EPR steering can be detected when
$\Delta >\varepsilon$. Fig.~\ref{LHS1}(a) shows the relation between
parameter $\theta$ and $\Delta$ obtained numerically (see the Methods section). We
find that $\Delta$ is of order $10^{-2}$ when $\theta$ is not closed
to $0$ or $\pi/2$. This implies that the EPR steering of state
$|\Psi\rangle_{AB}$ can be revealed by the experiments with
precision $\varepsilon<10^{-3}$. In Fig.~\ref{LHS1}(a), it is
observed that $\Delta$ changes symmetrically with respect to
$\theta$ and approaches to its maximal value when $\theta=\pi/4$.
This shows the more entangled the state is, the easier to detect EPR
steering in the experiment.

We next consider a two-qubit mixed state
\begin{eqnarray}
\rho_{AB}=\cos^2\theta|\psi^+\rangle_{AB}\langle\psi^+|+\sin^2\theta|\varphi^+\rangle_{AB}\langle\varphi^+|.
\label{rhoL}
\end{eqnarray}
Here
$|\psi^+\rangle_{AB}=\frac{1}{\sqrt{2}}(|00\rangle_{AB}+|11\rangle_{AB})$
and
$|\varphi^+\rangle_{AB}=\frac{1}{\sqrt{2}}(|01\rangle_{AB}+|10\rangle_{AB})$
are two Bell states.
The measurement settings of Alice are still $\hat{A} \in \{ \mathcal
{P}^{\hat{z}}_a, \mathcal {P}^{\hat{x}}_a \}$, and after Alice
performs measurement $\hat{A}$ and obtains result $a$, Bob's
conditional states can be expressed as
$\tilde{\rho}^{\hat{A},a}_B=\sum_\xi \wp(a|\hat{A},\xi) \wp_{\xi}
\rho_{\xi}$ provided with a LHS model $\{ \wp_{\xi} \rho_{\xi} \}$
and a stochastic map $\wp(a|\hat{A},\xi)$. Similarly, for the state
$\rho_{AB}$, Bob obtains quantum probabilities of measuring his
qubit in the states $|0\rangle_B$, $|1\rangle_B$, $|+\rangle_B$ and
$|-\rangle_B$ as
\begin{eqnarray}\label{NESF}
&&P^{\rm QM}_{1}=  {\rm Tr} [ |1\rangle_B \langle 1 | \tilde{\rho}^{\hat{z},0}_B ]=\frac{\sin^2\theta}{2}, \nonumber\\
&&P^{\rm QM}_{2}=  {\rm Tr} [ |0\rangle_B \langle 0 | \tilde{\rho}^{\hat{z},1}_B ]=\frac{\sin^2\theta}{2},\nonumber\\
&&P^{\rm QM}_{3}=  {\rm Tr} [ |-\rangle_B \langle - | \tilde{\rho}^{\hat{x},0}_B ]=0, \nonumber\\
&&P^{\rm QM}_{4}=  {\rm Tr} [ |+\rangle_B \langle + |
\tilde{\rho}^{\hat{x},1}_B ]=0,
\end{eqnarray}
where $| \pm \rangle_B = \frac{1}{\sqrt{2}}(|0\rangle_{B} \pm
|1\rangle_{B})$, and $\tilde{\rho}^{\hat{z},a}_B$ is Bob's
conditional state after Alice performs projective measurement in
$\hat{z}$-direction, etc. It has been proved that there does not
exist any LHS model for $\rho_{AB}$ with $\theta \neq
\pm\frac{\pi}{4}$~\cite{EPR-us} such that probability equations
in~(\ref{NESF}) can be satisfied simultaneously. If Bob observes
experimentally these four probabilities $P^{\rm QM}_{i}$'s, then EPR
steering of the state is exhibited, or there exists no LHS model.
Consider experimental imprecision and errors, we also investigate
the condition to detect EPR steering of $\rho_{AB}$ by plotting the
variation of LHS bound $\Delta$ versus $\theta$, see Fig.~\ref{LHS1}
(b). It can be found that for the experiments with precision
$\varepsilon<10^{-3}$, the EPR steering of $\rho_{AB}$ can be
observed when $\theta$ is not close to $\pi/4$. It is worthy of
pointing out that our test of EPR steering is not limited to the
states $|\Psi\rangle_{AB}$ and $\rho_{AB}$, but also applicable to
the family of two-qubit entangled states specified in Ref.
\cite{EPR-us}, regardless of pure or mixed.


\vspace{8pt}
\noindent{\bf Discussions}

\noindent Let us make some discussions
on the possible realization of our test in physical systems. We
first consider non-Abelian Fibonacci anyons which are shown to be
the simplest non-Abelian quasiparticles for universal topological
quantum computation~\cite{2008Nayak}. Follow Freedman \textit{et
al}.'s work ~\cite{2002Freedman}, we encode logical qubits into
triplets of anyons with total topological charge $1$:
$|0\rangle^L=|((\bullet,\bullet)_{\mathbf{I}},\bullet)_{\tau}\rangle$
and $|1\rangle^L=|((\bullet,\bullet)_{\tau},\bullet)_{\tau}\rangle$
(here $L$ denotes ``logical"). The so-called noncomputational state
$|\texttt{NC}\rangle=|((\bullet,\bullet)_{\tau},\bullet)_{\mathbf{I}}\rangle$
is the only state of three anyons that has total topological charge
$0$. Quantum operations can be constructed by using two elementary
braiding operations $R_1,R_2$ acting on the Hilbert space of three
Fibonacci anyons and their inverses~\cite{2005Bonesteel,2007Simon}. The obtained quantum gates, together with the controlled-NOT gate obtained in Refs.~\cite{2005Bonesteel,2007Simon,GHZ-us} are useful in the
construction of EPR steering test by preparing logical-qubit states and achieving required operations (see the Methods section).
Several candidates for realizing non-Abelian anyons have been suggested in physical systems,
such as fractional quantum Hall liquid~\cite{2004Xia}, rotating Bose-Einstein condensates~\cite{2001Cooper},
as well as quantum spin systems~\cite{2004Freedman,2005Fend}.

Another possible system to explore the realization of our test
experimentally is trapped ion.
Refs.~\cite{ion2006,ion20081,ion20082} have reported experimental
results of high-fidelity state preparation, quantum gate operations,
and state measurement for optical qubits stored in $^{40}{\rm
Ca}^{+}$ held in a trap. State preparation is usually done by
precisely manipulating the internal levels of ion utilizing laser
pulses and the Blatt group realized state initialization with
fidelity more than $99.8\%$~\cite{ion20081}. By a
M{\o}lmer-S{\o}rensen-type gate
operation~\cite{SMentangling1,SMentangling2}, a Bell-type entangled
state of ions with a fidelity of $99.3(1)\%$ was realized in the
same work~\cite{ion20081}. The Blatt group also presented
single-qubit gates with fidelity exceeding $99.9\%$ in trapped
ions~\cite{ion2006}. As for state measurement capability in an
ion-trap experiment, A. H. Myerson {\it et al.}~\cite{ion20082}
achieved $99.991(1)\%$ readout fidelity, sufficient for
fault-tolerant quantum computation by measuring population of states
using time-resolved photon counting. For the entangled state
realized in Ref.~\cite{ion20081}, the probability bound is found to
be $\Delta=0.0732$, and this means that the EPR steering of the
entangled state can be verified experimentally with precision
$\varepsilon<0.0732$. The experimental achievements in the
literatures~\cite{ion2006,ion20081,ion20082} tell us that our test
of EPR steering based on the AVN proof is possibly realizable with
current techniques in ion-trap experiments.

To summarize, we have presented a test to
identify EPR steering based on the AVN argument by measuring
projective probabilities. Our test is applicable to the family of
two-qubit entangled states specified in Ref.~\cite{EPR-us}
regardless of pure or mixed. We have provided the condition on
experimental implementation of our scheme through expression
(\ref{qqDT}) that EPR steering can be observed in the presence of
experimental imprecision and errors. Our result is the first
experimental test presented to detect EPR steering by resorting to
the AVN proof, and it can be implemented in systems such as
non-Abelian anyons and trapped ions. The primary advantage of our
test based on non-Abelian anyons is that it is fault-tolerant, or
the logical quantum state used is robust against local
perturbations. Specifically, it has been proven that these logical
qubits might be robust to random perturbations~\cite{2010}. Our test
can also be realizable in ion-trap experiments based on current
experimental techniques as recent progress in trapped ion offers
high-fidelity state preparation, quantum gate operations, and state
measurement for optical qubits stored in it. Let us point out that
the realization of our test is not limited to the two systems but
also applicable to many other physical systems like photons, atoms
and superconductors, etc. We expect further investigations in this
direction, both theoretically and experimentally.


\vspace{8pt}
\noindent{\bf Methods}

\noindent \textbf{To find the optimal LHS model.}
We here present a theorem which is used to find the optimal LHS model for a given two-qubit state.
\emph{Theorem -- For any given two-qubit state $\rho_{AB}$ in a $N$-setting protocol, if there is a LHS model for the state, then there is a LHS model with the number of hidden states no larger than $2^N$}.
The proof of the Theorem needs two lemmas associated with the concept of deterministic LHS (dLHS) model which is a LHS satisfying
 $\wp(a|\hat{A},\xi) \in \{0,1\}, \forall \hat{A}, \xi, a$. We also briefly restate the notations to be used, $\tilde{\rho}^{\hat{A}}_{a}$ is the conditional states of Bob after Alice measures $\hat{A}$ and gets result $a\in\{0,1\}$, the tilde here denotes  this state is unnormalized and its norm is $P^{\hat{A}}_{a}$, the probability associated with the output $a$.

\noindent \emph{Lemma 1.} For any given two-qubit state $\rho_{AB}$, if there is a LHS model for $\rho_{AB}$ then there is a dLHS model for $\rho_{AB}.$

In a general $N$-setting protocol, we have $\hat{A}\in\{\hat{A_1},\hat{A_2},\hat{A_3},...,\hat{A_N}\}$. Suppose $\rho_{AB}$ has a LHS description thus there is an ensemble $\{\wp_{\xi}\rho_{\xi}\}$ and an associated probability $\wp(a|\hat{A},\xi)$ fulfilling $\tilde{\rho}^{A}_a=\sum_{\xi} \wp(a|\hat{A},\xi) \wp_{\xi} \rho_{\xi}$.  We note that $\wp(1|\hat{A},\xi)=1-\wp(0|\hat{A},\xi)$. Now if $\forall \{\xi, \hat{A}\}, \wp(0|\hat{A},\xi)\in\{0,1\}$, then it is a dLHS model. We next check each $\xi$ to see whether $\wp(0|\hat{A},\xi)\in\{0,1\}$. For any $\xi$ with $\wp(0|\hat{A},\xi)\in\{0,1\}$, we keep these terms unchanged. For $\xi=k\;\mid$ $\wp(0|\hat{A},k)\notin\{0,1\}$, we decompose this term into $2^N$ separate terms as follows. First we define a new term $m_a=\sum_{i=1}^{N} 2^{N-i} a_i +1$, where $a_i$ denote the measurement results of $A_i$ ($a_i=0,1$). It is not difficult to find that $m_a$ ranges from $1$ to $2^N$ depending on $a_i$. We then do the decomposion by choosing
\begin{eqnarray}\label{dLHSM}
&&\rho_{k}^{(m_a)}= \rho_{k},\label{dLHSM1}\\
&&\wp_{k}^{(m_a)}=\prod_{i=1}^{N} \wp(a_i|\hat{A_i},k)\wp_k, \label{dLHSM2}\\
&&\wp(0|\hat{A_i},{k}^{(m_a)})= \left\{ \begin{aligned}
         &1 \;\;\; \textrm{if} \;\;a_i=0\\
         &0 \;\;\; \textrm{else}\end{aligned} \right.\label{dLHSM4}
\end{eqnarray}
where $\rho_{k}^{(m_a)}$ is the hidden state and $\wp_{k}^{(m_a)}$ is its weight. By direct calculations it can be verified that $\wp(a|\hat{A},k)\wp_{k}\rho_{k}=\sum^{2^N}_{m_a=1} \wp(a|\hat{A},{k}^{(m_a)}) \wp_{k}^{(m_a)} \rho_{k}^{(m_a)}$. Eq.~(\ref{dLHSM4}) shows the reconstructed stochastic maps are all deterministic. Thus by this way, we get a dLHS model that satisfies $\tilde{\rho}^{A}_a=\sum_{\xi} \wp(a|\hat{A},\xi) \wp_{\xi} \rho_{\xi}$. $\Box$

\noindent \emph{Lemma 2.} For a dLHS model, $\tilde{\rho}^{A}_a=\sum_{\xi} \wp(a|\hat{A},\xi) \wp_{\xi} \rho_{\xi}$ can be rewritten as $P^{\hat{A}}_{a} \rho^{\hat{A}}_{a}=\sum_{\xi \in H^{\hat{A}}_{a}} \wp_{\xi} \rho_{\xi}$, where $H^{\hat{A}}_{a}$ stands for the set of hidden states that contribute to $\rho^{{\hat{A}}}_{a}$ indicating the corresponding $\wp(a|\hat{A},\xi)=1$. The equality holds if and only if the following equalities are fulfilled,
\begin{equation}\label{MassCenter}
\left\{ \begin{aligned}
         P^{{\hat{A}}}_{a} &= \sum_{\xi \in H^{\hat{A}}_{a}} \wp_{\xi} \\
         P^{{\hat{A}}}_{a} \; \overrightarrow{r_{a}}^{{\hat{A}}} &= \sum_{\xi \in H^{\hat{A}}_{a}} \wp_{\xi} \; \overrightarrow{r_{\xi}}
        \end{aligned} \right.
\end{equation}
where $\overrightarrow{r_{a}}^{{\hat{A}}}$and $\overrightarrow{r_{\xi}}$ are the Bloch vectors of $\rho^{{\hat{A}}}_{a}$ and $\rho_{\xi}$ respectively.

Let us look at the proof of the lemma. We have $\rho^{{\hat{A}}}_{a} = (\mathbf{1}+\overrightarrow{r_{a}}^{{\hat{A}}} \cdot \overrightarrow{\sigma} )/2 $ and $\rho_{\xi} =(\mathbf{1}+\overrightarrow{r_{\xi}} \cdot \overrightarrow{\sigma} )/2$, where $\mathbf{1}$ describes identity matrix. So the equality of $P^{{\hat{A}}}_{a} \rho^{{\hat{A}}}_{a}=\sum_{\xi \in H^{\hat{A}}_{a}} \wp_{\xi} \rho_{\xi}$ gives $P^{{\hat{A}}}_{a} \, \mathbf{1} + P^{{\hat{A}}}_{a} \; \overrightarrow{r_{a}}^{{\hat{A}}} \cdot \overrightarrow{\sigma} =  \sum_{\xi \in H^{\hat{A}}_{a}} \wp_{\xi} \mathbf{1} + \sum_{\xi \in H^{\hat{A}}_{a}} \wp_{\xi} \overrightarrow{r_{\xi}} \cdot \overrightarrow{\sigma}$. Thus we obtain Eq.~(\ref{MassCenter}). $\Box$

We would like to point out that Eq. \eqref{MassCenter} is similar to the problem describing center of mass if we treat the probabilities $\wp_{\xi}$ and $P^{\hat{A}}_{a}$ as masses, as well as Bloch vectors ($\overrightarrow{r_{\xi}}$ and $\overrightarrow{r_{a}}^{{\hat{A}}}$) as the position vectors of various masses. Lemma 2 shows that the task to find a dLHS model for a state $\rho^{\hat{A}}_{a}$ with probability $P^{\hat{A}}_{a}$ is equivalent to find a distribution of masses in the Bloch sphere with total mass $P^{\hat{A}}_{a}$ and center of mass being located at $\overrightarrow{r_{a}}^{{\hat{A}}}$. We show in the following that with the aid of Eq. (\ref{MassCenter}), we can impose constraints on measurement settings to find a dLHS model. If we cannot find a dLHS model for $\rho_{AB}$, Lemma 1 shows that we can neither find a LHS model, and this thus affirms the steerability of $\rho_{AB}$. For any given $\rho_\xi$, we can always assign a $N$-length bit string constructed from $\wp(a|\hat{A_1},\xi)\wp(a|\hat{A_2},\xi)\cdots\wp(a|\hat{A_N},\xi)$ considering $\wp(a|\hat{A_i},\xi)\in\{0,1\}$. Next let us describe the LHS model by dividing hidden states $\{\rho_\xi\}$ into many subsets with each subset containing all of the $\rho_\xi$ that has the same $N$-length bit string. Thus in this way, each subset is unique, or not overlapping with others. We can take each of the subsets as one new hidden state by resorting to Lemma 2. We use the fact that hidden state can be treated as mass point so we can consider the centre of mass of each subset as the new state and the weight of the new state is the correspoinding total mass. It is not difficult to find that there are totally $2^N$ such new states, and thus the LHS model has only $2^N$ hidden states. This ends our proof of the Theorem.

\begin{figure}
\includegraphics[width=150mm]{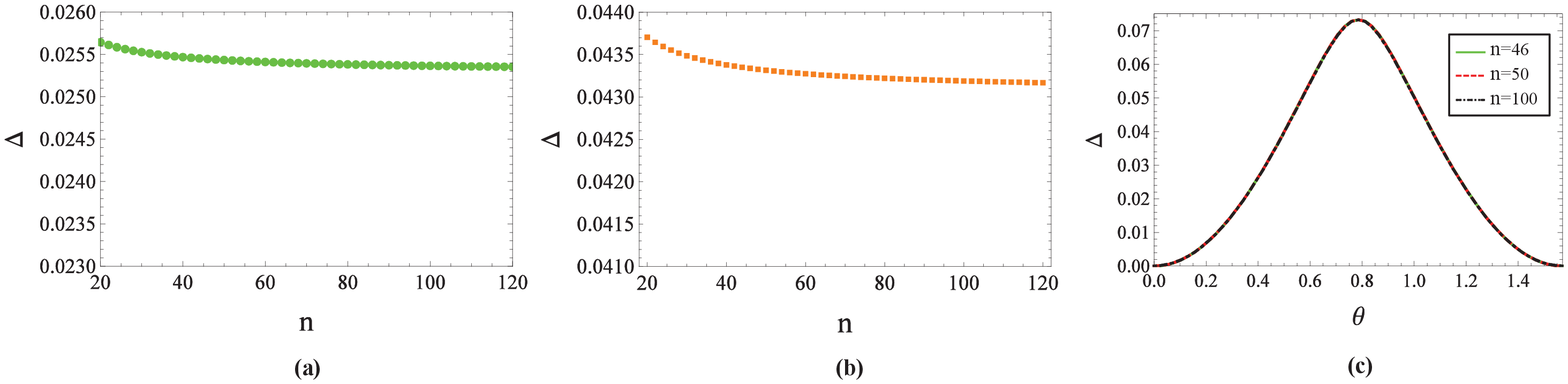}\\
\caption{\textbf{Variations of $\Delta$ for $|\Psi\rangle_{AB}$ versus $n$/$\theta$.} (a) $\theta=\pi/8$ and $n$ ranges from $20$ to $120$, (b) $\theta=\pi/6$ and $n$ ranges from $20$ to $120$, as well as (c) $n=46,50,100$ and $\theta$ ranges from $0$ to $\pi/2$. From (a) and (b), we find that the variation of $\Delta$ is negligibly small when $n>45$ since the variation is in the ten-thousandths place. From (c) it is clear that the three curves are almost overlapped and the results show that $n=46$ is large enough to obtain a reasonable value of $\Delta$.}\label{Dnum}
\end{figure}

Therefore we conclude that the optimal LHS model contains an
ensemble with four pure hidden states in the two-setting protocol, and more hidden states make no improvement.
The optimal LHS model is numerically obtained by minimizing
function $ F_{n}= \Sigma_{i=1}^{4} v_i^n $ for a large $n$, where $v_i=|P^{\rm LHS}_{i}-P^{\rm
QM}_{i}|$. In this approach, we utilize the knowledge of vector norm. First we have $(F_{n})^{1/n}$ which is the $l_n$-norm of vector $\vec{v}=(v_1, v_2, v_3, v_4)^{T}$. We know that $l_\infty$-norm of a vector is just its
maximum element and hence by definition $\Delta$ equals to the minimum of
$l_\infty$-norm of the vector $\vec{v}=(v_1, v_2, v_3, v_4)^{T}$. So for a large enough
$n$ we can get a good approximation of $\Delta$ from minimizing $(F_{n})^{1/n}$ with varied $\rho_\xi$. In our calculations, we use $n=46$ since we find numerically the improvement of $\Delta$ by choosing a number larger than $46$ is negligibly small. As shown in Fig. \ref{Dnum} (a) and (b), we find the values of $\Delta$ by choosing $\theta=\pi/8,\;\pi/6$ for $|\Psi\rangle_{AB}$ with different $n$ (ranging from $20$ to $120$). It is clear that the values of $\Delta$ do not change substantially and the change is in the ten-thousandths place when $n$ is greater than $45$. The results show us that $n=46$ is large enough to obtain a reasonable value of $\Delta$. Apparemently we can choose other values of $n$ as long as $n>45$ and the choice will not affect the value of $\Delta$ much. We also plot the variation of $\Delta$ versus $\theta$ by choosing different $n$ for $|\Psi\rangle_{AB}$ in Fig. \ref{Dnum} (c). Seen from Fig. \ref{Dnum} (c), the three curves corresponding to $n=46$, $n=50$, $n=100$ respectively are almost overlapped. Hence we know that $n=46$ is large enough to obtain a reasonable value of $\Delta$.

\vspace{6pt}

\begin{figure}
\includegraphics[width=80mm]{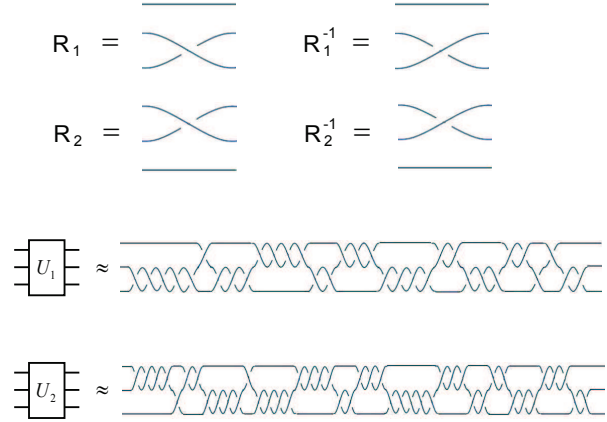}\\
\caption{ \textbf{Approximating quantum gates $U_{\theta}$ by
braiding Fibonacci anyons.} In the plotting, time flows from left to
right, $U_1$ represents $U_{\pi/6}$ and $U_2$ represents
$U_{-\pi/3}$.}\label{Braid-Gates}
\end{figure}

\noindent\textbf{Approximation of quantum gates in non-Abelian
Fibonacci anyons.} Quantum operations can be constructed by using
two elementary braiding operations $R_1,R_2$ acting on the Hilbert
space of three Fibonacci anyons and their
inverses~\cite{2005Bonesteel,2007Simon}. In Fig.~\ref{Braid-Gates},
we plot the braids that approximate the quantum gate
\begin{eqnarray}\label{Ut}
U_{\theta}=\left(\begin{matrix}
\cos\theta&\sin\theta\\
\sin\theta&-\cos\theta
\end{matrix}\right)
\end{eqnarray}
with $U_1=U_{\pi/6}$ and $U_2=U_{-\pi/3}$. Any other quantum gate
$U_{\theta}$ can be obtained in a similar way.
The approximations are obtained by performing brute force searches and
the distance between two matrices $\mathscr{M}$ and
$\mathscr{M}^{\prime}$ is defined as the square root of the largest
eigenvalues of $(\mathscr{M}-\mathscr{M}^{\prime})^{\dagger}
(\mathscr{M}-\mathscr{M}^{\prime})$~\cite{2005Bonesteel,2007Simon}. The
distances between the required operations and the gates resulting
from actual braiding are about $5.7\times10^{-5}$ for $U_1$ and
$U_2$. In fact, these gates can be systematically improved to any
required accuracy due to the Solovay-Kitaev
theorem~\cite{1999Kitaev}. The above quantum gates, together with the controlled-NOT gate
obtained in Refs.~\cite{2005Bonesteel,2007Simon,GHZ-us} are useful in the
construction of EPR steering test by preparing
logical-qubit states. We apply the
operation $U_{\theta}$ (with $\theta\in (0, \pi/2)$) on the logical
qubit A of initial state $|\Psi\rangle^L_0=|0\rangle^L_A\otimes|0\rangle^L_B$ and a controlled-NOT gate is followed on the two logical
qubits, then we have the two-logical-qubit pure states $|\Psi\rangle^L_{AB}$.
To prepare mixed state, we need an ancilla logical qubit C, and
initially assume that the logical qubits are in the state
$|\Psi\rangle^L_0=|0\rangle^L_A\otimes|0\rangle^L_B\otimes|0\rangle^L_C$.
We apply Hadamard gate on logical qubit A, $U_{\theta}$ on logical qubit C, then a controlled-NOT gate on logical qubits A and
B, and finally a controlled-NOT gate on logical qubits C and B, we
then have
$|\Psi\rangle^L_{ABC}=\cos\theta|\psi^+\rangle^L_{AB}|0\rangle^L_C+\sin\theta|\varphi^+\rangle^L_{AB}|1\rangle^L_C$.
Look at the first two qubits only, we successfully have the
state $\rho^L_{AB}$ as in (\ref{rhoL}).
All the operations involved in our scheme, such as $\mathcal {P}^{\hat{z}}_a$
and $\mathcal {P}^{\hat{x}}_a$ for Alice, $|\psi^\perp \rangle^L_B
\langle \psi^\perp |$ and $|\varphi^\perp \rangle^L_B \langle
\varphi^\perp | $ for Bob, can be carried out by braiding the
Fibonacci anyons. For instance, the single-logical-qubit states
$|\psi^\perp\rangle^L_B$ and $|\varphi^\perp\rangle^L_B$ of Bob can
be realized by using the action of $U_{\pm\theta}$ on
$|1\rangle^L_B$ (up to a global phase).


{\bf Acknowledgements}

 We thank V. Vedral for valuable discussions.
J.L.C. is supported by National Basic Research Program (973 Program)
of China under Grant No. 2012CB921900 and NSF of China (Grant No.
11175089). This work is also partly supported by National Research
Foundation and Ministry of Education, Singapore (Grant No. WBS:
R-710-000-008-271).

{\bf Author Contributions}

 All authors contributed to this work including developing the scheme and preparing the manuscript.
 C.W. and J.L.C. initiated the idea. C.W., J.L.C., X.J.Y., H.Y.S., and D.L.D. proposed the scheme.
C.W., J.L.C., Z.W. and C.H.O. wrote the main manuscript text. X.J.Y.
and H.Y.S. prepared the figures. All authors reviewed the
manuscript.

{\bf Supplementary Information} is linked to the online version of
the paper at www.nature.com/nature.

{\bf Additional Information}

{\bf Competing financial interests:} The authors declare no
competing financial interests.

\newpage

\end{document}